\documentstyle[prl,aps,epsf,psfig,rotating,graphics]{revtex}
\newcommand \be{\begin{eqnarray}}
\newcommand \ee{\end{eqnarray}}
\begin{document}
\twocolumn[\hsize\textwidth\columnwidth\hsize
           \csname @twocolumnfalse\endcsname
\title{Space-time versus particle-hole symmetry in quantum
Enskog equations}
\author{V\'aclav \v Spi\v cka$^{1,2}$ and Klaus Morawetz$^{2}$
and Pavel Lipavsk\'y$^{1,2}$}
\address{$^{1}$Institute of Physics, Academy of Sciences,
Cukrovarnick\'a 10, 16200 Praha 6, Czech Republic\\
$^{2}$Max Planck Institute for the Physics of Complex Systems,
N\"othnitzer Str. 38, 01187 Dresden, Germany }
\maketitle
\begin{abstract}
The non-local scattering-in and -out integrals of the Enskog equation
have reversed displacements of colliding particles reflecting that the
-in and -out processes are conjugated by the space and time inversions.
Generalisations of the Enskog equation to Fermi liquid systems are
hindered by a request of the particle-hole symmetry which contradicts
the reversed displacements. We resolve this problem with the help of
the optical theorem. It is found that space-time and particle-hole
symmetry can only be fulfilled simultaneously for the Bruckner-type
of internal Pauli-blocking while the Feynman-Galitskii form allows
only for particle-hole symmetry but not for space-time symmetry due to
a stimulated emission of Bosons.
\end{abstract}
\vskip2pc]

\section{Introduction}
The formulation of kinetic theory of dense interacting Fermi gases
beyond the Boltzmann equation (BE) is an ongoing task. For classical
hard-sphere gas the main theoretical focus was on the statistical
correlations resulting in the Enskog equation
\cite{E21,CC90,C62,W65,KO65,DC67,BE79}. In contrast to the BE, the
collision integral of the Enskog equation is non-local what takes
into account that when two hard spheres collide, their centres are
displaced by the sum of their radii. The particle scattered out of
its free trajectory faces its collision partner in front, while the
particle scattered in the new free trajectory leaves its partner
behind.  This is expressed by the opposite signs of non-local
corrections in the scattering-out and -in integrals.

Various generalisations of the Enskog equation towards quantum
systems \cite{B69,La89,L90,SN90,H90,NTL91,BKKS96} have been
developed mostly in the last two decades. They offer numerous gradient
corrections to the scattering integral which describe how the
non-local character of collisions contributes to smooth
perturbations. With a typical number of gradient corrections counted
in tens, a comparison of the original Enskog equation with its
generalisations was not possible.

The connection became more clear after Laloe, Nacher and Tastevin
\cite{NTL91} recognized that some of obtained gradient corrections
can be recast into effective fields and renormalizations of the
mass of particles, i.e., these gradient corrections are linked to
the Landau concept of quasiparticles. They also show that when the
quasiparticle contributions are separated, all remaining gradient
contributions are proportional to various derivatives of the
scattering phase shift. These derivatives have natural link to
the Wigner collision delay \cite{DP96} which also describes the
non-locality of collisions, although in the time not in the space.

A kinetic equation which combines the non-locality in time and
space has been derived as the quasi-classical asymptotics of
non-equilibrium Green's functions \cite{SLM98,Mt98}. In
\cite{SLM98} a backward resummation of the gradient expansion was
introduced by which one obtains the scattering integral in a form
reminding the Enskog equation: the gradient corrections are
expressed as shifts of arguments in the initial (final) condition
so that one can see how long the collision lasts and how far from
each other are particles at the beginning (end) of a collision. Of
course, the hard-sphere gas is a special case to which the theory
applies. It turns out that the scattering-in is identical to the
Enskog equation while the displacement of the scattering-out does
not have the expected opposite sign. A careful inspection shows
that this sign problem appears also within all earlier approaches
\cite{B69,La89,L90,SN90,H90,NTL91,BKKS96}.

The sign puzzle has two serious consequences for the applicability
of the non-local kinetic equation. First, the Enskog equation
corresponds to classical trajectories, therefore it can be
numerically studied either with the Monte-Carlo simulation or by
the so called molecular dynamics. The kinetic equations derived
from the quantum statistics cannot be studied with these methods.
Second, the Enskog equation yields the hydrodynamic Chapman-Enskog
expansion in a straight and relatively simple manner \cite{CC90}
what allows one to identify the thermodynamic properties of the
system. The symmetry between the scattering-in and -out is a very
important prerequisite in separation of cancelling and conserving
quantities. Without this symmetry, one can also derive conservation
laws \cite{H90}, however, an extensive application of physically
non-transparent identities is necessary.

In this paper we show how the natural symmetry of the Enskog equation
can be obtained within the quantum mechanical approach to the kinetic
equation. In the next section we introduce the problem of symmetry in
a naive manner using {\em ad hoc} kinetic equations for the Fermi
liquid. In Section~III we show that the non-local corrections for the
Fermi liquid are linked to in-medium effects and provide an identity
which allows one to achieve the Enskog form of non-local corrections.
Section~IV includes conclusions.

\section{Classical versus quantum collision}
The problem with sign in the scattering-out follows from a difference
between the classical and quantum approaches to collisions. One has to
recognize that a realistic collision has a finite duration $\Delta_t$
and to compare these two approaches in the time picture.

\subsection{Pseudo-classical approach}
The simplest model system on which one can illustrate both approaches
is a homogeneous gas of particles which form short-living molecules,
i.e., the system with a dominant resonant scattering \cite{DP96}.
Its kinetic equation reads
\begin{equation}
{\partial f_k(t)\over\partial t}=
\int PF_{k+p}(t)-\int P f_k(t) f_p(t).
\label{1ONEMOL}
\end{equation}
The last term is the scattering-out which describes that with the
probability $P$ two particles form a molecule and thus a particles
leaves the state of momentum $k$. The first term on the right hand
side corresponds to the decay of the molecule into two particles,
one of them achieves momentum $k$. The dependence of the distribution
of molecules $F$ is also covered by the balance equation,
\begin{equation}
{\partial F_K(t)\over\partial t}=
\int P f_{K-p}(t)f_p(t)-{F_K(t)\over\Delta_t}.
\label{1TWOMOL}
\end{equation}
The last term describes the decay of molecules with lifetime
$\Delta_t$, the first term on the right hand side their formation.

The balance equation (\ref{1TWOMOL}) for molecules is solved by
\begin{eqnarray}
F_K(t)&=&\int\limits^{\infty}_0 d\tau {\rm e}^{-{\tau\over\Delta_t}}
\int P f_{K-p}(t-\tau)f_p(t-\tau)
\nonumber\\
&\approx& \int P f_{K-p}(t-\Delta_t)f_p(t-\Delta_t).
\label{Flin}
\end{eqnarray}
The second line is the gradient approximation which is sufficient
for our discussion since all quantum approaches to the non-local
kinetic equation are restricted to it. Using (\ref{Flin}) in
(\ref{1ONEMOL}) one gets a kinetic equation,
\begin{equation}
{\partial f_k(t)\over\partial t}=
\int Pf_{k-q}(t\!-\!\Delta t)f_{p+q}(t\!-\!\Delta_t)
-\int Pf_k(t)f_p(t).
\label{1FIVEMOL}
\end{equation}
The scattering-in has a retarded initial condition reflecting that
the molecule lives from $t-\Delta_t$ to $t$. The initial condition
of the scattering-out is associated with time instant $t$, the
corresponding molecule thus lives from $t$ to $t+\Delta_t$.

In dense Fermi systems, the final states of collisions might be
occupied and the collision is then prohibited. Let us modify
kinetic equation (\ref{1FIVEMOL}) by {\em ad hoc} Pauli blocking
factors as introduced by Nordheim, Uehling and Uhlenbeck \cite{N28,UU33}
\begin{eqnarray}
{\partial f_k(t)\over\partial t}&=&
\int Pf_{k-q}(t-\Delta_t)f_{p+q}(t-\Delta_t)
\nonumber\\
&&\times\left(1-f_k(t)\right)\left(1-f_p(t)\right)
\nonumber\\
&-&\int Pf_k(t)f_p(t)
\nonumber\\
&&\times\left(1-f_{k-q}(t+\Delta_t)\right)
\left(1-f_{p+q}(t+\Delta_t)\right).
\label{Fbloc}
\end{eqnarray}
The time arguments of the blocking factors, $1-f$, correspond to ends
of time intervals during which the collision happens because the
blocking is attributed to the final states.

\subsection{Quantum approach}
One can see that the scattering-out of (\ref{Fbloc}), obtained within
pseudo-classical assumptions, requires the blocking factor at future
time $t+\Delta_t$. The quantum approach, however, does not allow to
look into future and treats the same
process differently.

\begin{figure}[h]
\scalebox{0.7}{\includegraphics*[10mm,10mm][120mm,93mm]{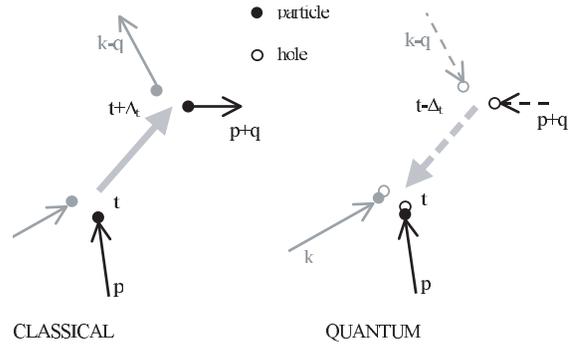}}
\caption{Scattering-out for classical and quantum concept of
collisions.}
\label{f3}
\end{figure}

In the quantum statistics, the scattering-out is described as a
collision of two holes, see figure~\ref{f3}. In our simple model,
two holes form a hole-molecule which also lives for $\Delta_t$. When
this hole-molecule decays into two holes, these holes annihilate
particles of corresponding  momenta. Accordingly, the scattering-out
is described by the hole-hole interaction during the time interval
from $t-\Delta_t$ to $t$. An {\em ad hoc} kinetic equation
corresponding to the quantum picture thus reads
\begin{eqnarray}
{\partial f_k(t)\over\partial t}&=&
\int Pf_{k-q}(t-\Delta_t)f_{p+q}(t-\Delta_t)
\nonumber\\
&&\times\left(1-f_k(t)\right)\left(1-f_p(t)\right)
\nonumber\\
&-&\int Pf_k(t)f_p(t)
\nonumber\\
&&\times\left(1-f_{k-q}(t-\Delta_t)\right)
\left(1-f_{p+q}(t-\Delta_t)\right).
\label{Fhole}
\end{eqnarray}
Note that (\ref{Fhole}) differs from its pseudo-classical counterpart
(\ref{Fbloc}) by the sign of the non-local correction. This is the
time-modification of the sign problem found for the quantum
generalisations of the Enskog equation.

The above {\em ad hoc} implementations, Eq.'s (\ref{Fbloc}) and
(\ref{Fhole}), of the non-local
corrections reveal a paradox: The space-time symmetry and the
particle-hole symmetry lead to contradictory results. Indeed,
equations (\ref{Fbloc}) and (\ref{Fhole}) are different and for a
general scattering rate $P$ they correspond to different thermodynamic
properties of the system.

\section{In-medium effects}
To resolve the paradox of symmetries, one has to take into account
that the scattering rate $P$ itself is a function of the occupation,
$P[f]$. This dependence represents an internal Pauli blocking of
states during collisions,
which is called  in-medium effect in nuclear physics.
In a heuristic manner one can indicate what kind of internal Pauli
blocking is consistent with the Uehling-Uhlenbeck blocking of
final states.

Since the scattering process lasts over the time interval from
$t-\Delta_t$ to $t$, the mean value of $P$ equals its value at the
centre time $P=P(t-{1\over 2}\Delta_t)$, see \cite{SLM98,Mt98}.
Comparing equations (\ref{Fbloc}) and (\ref{Fhole}) we find that
the two symmetries are consistent if
\begin{eqnarray}
&&\int P\left(t-{1\over 2}\Delta_t\right)f_k(t)f_p(t)
\nonumber\\
&&~~~~~\times\left(1-f_{k-q}(t-\Delta_t)\right)
\left(1-f_{p+q}(t-\Delta_t)\right)
\nonumber\\
&=&\int P\left(t+{1\over 2}\Delta_t\right)f_k(t)f_p(t)
\nonumber\\
&&~~~~~\times\left(1-f_{k-q}(t+\Delta_t)\right)
\left(1-f_{p+q}(t+\Delta_t)\right).
\label{symeq}
\end{eqnarray}
Within the gradient approximation this condition reads
\begin{eqnarray}
&&\int P f_kf_p(1-f_{k-q})(1-f_{k-q})
\nonumber\\
&&\times\left({d\ln P\over dt}+2{d\over dt}
\ln [(1-f_{k-q})(1-f_{p+q})]\right)=0.
\label{symeqlin}
\end{eqnarray}
>From this equation we see that the space-time and particle-hole
symmetries are consistent when the time dependence of in-medium effect
is given by the
Pauli blocking of the Bruckner type, $P[(1-f)(1-f)]$. According to
this type of the Pauli blocking, the internal states of the
short-living molecule exist only in the unoccupied phase space.

\subsection{Causality}
The retarded scattering-out integral of Enskog type (\ref{Fbloc})
is peculiar from the point of view of the causality. Since the
scattering-out process ends at time $t+\Delta_t$, the scattered
particles have to have available final state at this time. In other
words, to determine whether the collision is allowed by the Pauli
exclusion principle, one has to look into future. In such a way,
the Pauli blocking seems to bring an anti-casual step.

In general, the causality of the perturbative expansion reflects 
the tendency of a many-body system to reach its equilibrium state. 
The anti-causal descriptions of the whole system is thus impossible
because of the dissipative processes. Accordingly, we will take the
causal expansion and the subsequent the particle-hole symmetry 
represented by (\ref{Fhole}) as well justified starting point.

The Enskog-type kinetic equation with the space-time symmetry of 
the scattering integral applies only under restrictive assumptions.
The first assumption is that individual binary collisions are 
treated as if they were isolated from the rest of the system.
The dynamics of the binary collision is then reversible and the 
causal and anti-causal expansions on the space-time scale of a
single collision are equivalent. This assumption is met in all
approaches to the kinetic equation except for the studies of the 
so called collisional broadening. The second assumption is that
the internal Pauli blocking of collisions is of the Bruckner type.
This point is discussed bellow.

All methods of the quantum statistics enforce the causality using
backward propagation of holes instead of the forward propagation
of particles into future. To link the space-time and particle-hole
symmetries, we will use the optical theorem which allows us to
reformulate the causal internal propagation during the collision
into the anti-causal one.

In the algebraic notation of the double-time Green
functions\cite{Kel64,LW72,BS94}, the causality is reflected
by the order of operators, `retarded--correlation--advanced'. The time
cuts of the retarded and advanced operators restrict all time integrals
to the past. The anti-causal expansion is then characterised by the
reversed order, `advanced--correlation--retarded'. Without
introducing unnecessary details, we can link the causal and
anti-causal expansions using the identity for the scattering T-matrix,
\begin{equation}
T^R{\cal A}T^A=T^A{\cal A}T^R.
\label{e8}
\end{equation}
This identity represents two forms of the optical theorem, ${\rm Im}
T=T^R{\cal A}T^A$ and ${\rm Im}T=T^A{\cal A}T^R$. Their derivations
are enclosed in Appendix~A.

The retarded/advanced T-matrix, $T^{R,A}$, describes an individual
binary process \cite{GW64}. The two-particle spectral function
$\cal A$ includes the internal Pauli blocking. In this paper we
discuss two particular approximations of the internal Pauli blocking, 
the Bruckner approximation,
\begin{equation}
{\cal A}_B(t_1,t_2,k,p)\approx (1-f_k)(1-f_p){\rm e}^{-i(\epsilon_k+
\epsilon_p)(t_1-t_2)},
\label{ABruc}
\end{equation}
and the Galitskii-Feynman approximation,
\begin{equation}
{\cal A}_{GF}(t_1,t_2,k,p)\approx (1-f_k-f_p){\rm e}^{-i(\epsilon_k+
\epsilon_p)(t_1-t_2)}.
\label{AGalF}
\end{equation}
For both approximations we will derive kinetic equations with the
space-time symmetry of the non-local scattering integral. We will 
see that the kinetic equation obtained within the Bruckner approximation 
has the pseudo-classical form (\ref{Fbloc}), therefore it can be
treated with numerical tools based on classical concept of trajectories.
In contrast, the kinetic equation within the Galitskii-Feynman 
approximation includes a non-trivial term due to the stimulated emission 
of Bosons which essentially complicates its numerical treatment.

\subsection{Collision integral from Green functions}
Let us first remind how the non-local scattering integrals relate to
more general relations of the quantum statistics. We demonstrate it
on the method of non-equilibrium Green's functions. The scattering-in
and \mbox{-out} integrals result from anti-commutators, $\{.,.\}$, of 
the Kadanoff and Baym (KB) equation \cite{KB62,D84,BM90}
\begin{eqnarray}
\{G^>,\Sigma^<\}\!-\!\{G^<,\Sigma^>\}&=&
\left\{G^>,G^>\circ T^R(G^<G^<)T^A\right\}
\nonumber\\
&-&\left\{G^<,G^<\circ T^R(G^>G^>)T^A\right\}.
\nonumber\\
\label{e7}
\end{eqnarray}
Here, $G^<$ and $G^>$ are particle and hole correlation functions,
the $\circ$ denotes that $G^{>,<}$ closes one loop of the two-particle
function on its right hand side. The T-matrices and pairs of
single-particle correlation functions, $(GG)$, obey standard
two-particle operator products.

With respect to our treatment, it is sufficient to know that the
particle correlation functions are proportional to the quasiparticle
distribution, $G^<(t_1,t_2,k)\approx f_k{\rm
e}^{-i\epsilon_k(t_1-t_2)}$,
therefore they represent the initial states of collisions. Similarly,
the hole correlation functions are proportional to the hole
distribution,
$G^>(t_1,t_2,k)\approx (1-f_k){\rm e}^{-i\epsilon_k(t_1-t_2)}$,
therefore they describe the final states including their Pauli blocking.

According to initial and final states, the first and second terms of the
right hand side of (\ref{e7}) can be interpreted as the scattering-in
and \mbox{-out} integrals, respectively. Note that both scattering 
integrals are causal having order `retarded--correlation--advanced' 
of the two-particle functions. In the same time, the scattering-in 
and \mbox{-out} integrals are linked via the particle-hole symmetry. 
One can see that upon the interchange of particles and holes, 
$>\longleftrightarrow <$, the first term changes to the second one and 
vice versa. Equation (\ref{e7}) is thus a precursor of (\ref{Fhole}).

Using the (extended) quasiparticle and quasi-classical approximations
keeping gradients in the scattering integral of the KB equation, one
obtains a non-local kinetic equation \cite{SLM98},
\begin{eqnarray}
{\partial f_1\over\partial t}&+&{\partial\epsilon_1\over\partial k}
{\partial f_1\over\partial r}-{\partial\epsilon_1\over\partial r}
{\partial f_1\over\partial k}
\nonumber\\
&=&\int P^-\ f_3^-f_4^-\bigl(1-f_1\bigr)\bigl(1-f_2^-\bigr)
\nonumber\\
&-&\int P^-\ \bigl(1-f_3^-\bigr)\bigl(1-f_4^-\bigr)f_1f_2^-.
\label{e11ni}
\end{eqnarray}
Algebraic operations needed to arrive at (\ref{e11ni}) are rather
extensive due to numerous gradient contributions to the scattering
integrals. These gradient contributions are expressed via shifts of
arguments as
\begin{eqnarray}
f_1&\equiv&f(k,r,t)
\nonumber\\
f_2^-&\equiv&f(p,r\!-\!\Delta_2,t)
\nonumber\\
f_3^-&\equiv&f(k\!-\!q\!-\!\Delta_K,r\!-\!\Delta_3,t\!-\!\Delta_t)
\nonumber\\
f_4^-&\equiv&f(p\!+\!q\!-\!\Delta_K,r\!-\!\Delta_4,t\!-\!\Delta_t).
\label{fsh}
\end{eqnarray}
The differential cross section is proportional to the square of the
amplitude of the T-matrix
\begin{eqnarray}
&&P^-={dp\over(2\pi)^3}{dq\over(2\pi)^3}2\pi\delta\left(\epsilon_1+
\epsilon_2^--\epsilon_3^--\epsilon_4^--2\Delta_E\right)
\nonumber\\
&&\left|T\!\left(\!\epsilon_1\!+\!\epsilon_2^-\!-\!\Delta_E,k-{\Delta_K
\over 2},p-{\Delta_K\over 2},q,r-\Delta_r,t-{\Delta_t\over 2}\right)
\right|^2\!.
\nonumber\\
\label{Psh}
\end{eqnarray}
Arguments of quasiparticle energies, $\epsilon$'s, are identical
with (\ref{fsh}).
All non-local corrections are given by derivatives of the scattering
phase shift \mbox{$\phi={\rm Im\ ln}T^R(\Omega,k,p,q,t,r)$}
\cite{SLM98},
\begin{eqnarray}
\Delta_t&=&{\partial\phi\over\partial\Omega},\ \ \
\Delta_E=-{1\over 2}{\partial\phi\over\partial t},\ \ \
\Delta_K={1\over 2}{\partial\phi\over\partial r},\ \ \
\Delta_3=-{\partial\phi\over\partial k},
\nonumber\\
\Delta_2&=&{\partial\phi\over\partial p}-{\partial\phi\over\partial q}-
{\partial\phi\over\partial k},\ \ \ \ \ \ \ \ \ \
\Delta_4=-{\partial\phi\over\partial k}-{\partial\phi\over\partial q}.
\label{e6}
\end{eqnarray}

A detail understanding of these numerous corrections to the scattering
integral of the Boltzmann equation is not essential for our discussion.
It is important to realize that the collision is of finite duration
$\Delta_t$. During this time particles can gain momentum and energy,
$\Delta_{K,E}$ due to the medium effect on the collision. Three
displacements $\Delta_{2,3,4}$ correspond to initial and final
positions of two colliding particles/holes.

The quasiparticle kinetic equation (\ref{e11ni}) covers three
ingredients of the kinetic theory. First, the scattering integral
includes medium effect on the scattering rate. Second, the scattering
integrals are non-local in space and time. Third, the quasiparticle
energy represents the momentum dependent mean field. With respect to
included non-local corrections, it is important that the quasiparticle
energy is defined from the pole of the propagator, not from the
variation of the energy density. This difference has been discussed
in \onlinecite{LSM99}. The scattering-out of (\ref{e11ni}) is the
particle-hole mirror of the scattering-in, accordingly it is not
the space-time mirror found in the Enskog equation.

\subsection{Anti-causal collision integral}
Our aim is to rearrange (\ref{e11ni}) so that it will include the
scattering-out as the space-time mirror of the scattering-in, briefly,
it will be the symmetry assumed by Enskog. It is advantageous to
make this step already on the level of Green's functions. Accordingly,
we rearrange (\ref{e7}) so that its scattering-out part is written in
terms of the anti-causal expansion.

Further progress depends on the approximation of the T-matrix. Let us
first approximate the T-matrix by Bruckner's reaction matrix for which 
the two-particle spectral function is ${\cal A}_B=(G^>G^>)$. Formula
(\ref{ABruc}) is the quasiparticle approximation of $(G^>G^>)$. Using 
identity (\ref{e8}) in the second term of (\ref{e7}) one finds
\begin{eqnarray}
\{G^>,\Sigma^<\}\!-\!\{G^<,\Sigma^>\}&=&
\left\{G^>,G^>\circ T^R(G^<G^<)T^A\right\}
\nonumber\\
&-&\left\{G^<,G^<\circ T^A(G^>G^>)T^R\right\}.
\nonumber\\
\label{e9}
\end{eqnarray}

Expression (\ref{e9}) has the desired explicit space-time symmetry in
contrary to explicit particle-hole symmetry of (\ref{e7}). To see it
in detail, we use (\ref{e9}) in the KB equation and employ the same
steps as above (quasi-classical and quasiparticle approximations) to
arrive at
\begin{eqnarray}
{\partial f_1\over\partial t}&+&{\partial\epsilon_1\over\partial k}
{\partial f_1\over\partial r}-{\partial\epsilon_1\over\partial r}
{\partial f_1\over\partial k}
\nonumber\\
&=&\int P^-\ f_3^-f_4^-\bigl(1-f_1\bigr)\bigl(1-f_2^-\bigr)
\nonumber\\
&-&\int P^+\ \bigl(1-f_3^+\bigr)\bigl(1-f_4^+\bigr)f_1f_2^+.
\label{e11i}
\end{eqnarray}
The shifts in the scattering-out have opposite signs,
\begin{eqnarray}
f_2^+&\equiv&f(p,r\!+\!\Delta_2,t)
\nonumber\\
f_3^+&\equiv&f(k\!-\!q\!+\!\Delta_K,r\!+\!\Delta_3,t\!+\!\Delta_t)
\nonumber\\
f_4^+&\equiv&f(p\!+\!q\!+\!\Delta_K,r\!+\!\Delta_4,t\!+\!\Delta_t),
\label{fshpl}
\end{eqnarray}
and similarly other ingredients denoted by superscript $+$.

It can be indicated why the change of the causal picture into the
anti-causal one results in the flipped signs of all non-local
corrections. First, one can use a formal argument. Writing the
T-matrices as products of the amplitude and the phase,
$T^R=|T|{\rm e}^{i\phi}$ and $T^A=|T|{\rm e}^{-i\phi}$, one can
see that the interchange of retarded and advanced T-matrices merely
flips the sign of the phase shift $\phi$. As all $\Delta$'s depend
linearly on $\phi$, the gradient contributions of the anti-causal
scattering-out have signs reversed with respect to the causal
scattering-in. Second, there is a physical reason for the formal
argument above. The amplitude of T-matrix represents a filter which
selects a probability of individual channels. The factor of phase shift,
${\rm e}^{i\phi}$, is a unitary transformation which applies to
individual components of the wave function in a manner which parallels
the evolution operator. Products like ${\rm e}^{i\phi}\ldots
{\rm e}^{-i\phi}$ correspond to transformation from one place to
another, while ${\rm e}^{-i\phi}\ldots{\rm e}^{i\phi}$ to the
backward one.

For the system of classical hard spheres, kinetic equation
(\ref{e11i}) reduces to the Enskog equation in the second order
virial approximation. This limit includes three simplifications.
First, the Pauli blocking factors vanish in the classical limit,
$1-f\to 1$. Second, the quasiparticle energy reduces to the kinetic
energy of free particles, $\epsilon_1\to {k^2\over 2m}$. For this
limit it is important that the quasiparticle energy is defined from
the pole of the propagator. Landau's definition of $\epsilon$ based
on the variation of the energy density yields a non-trivial
quasiparticle energy even for the classical gas of hard spheres.
Third, from the hard-sphere scattering phase shift, $\phi\to\pi-|q|D$,
where $D$ is a diameter of colliding particles, one finds expected
values of $\Delta$'s. The collision delay is zero, $\Delta_t\to 0$,
there is no energy/momentum gain, $\Delta_{E,K}\to 0$, during
collision none of particles move in space, $\Delta_3=0$ and
$\Delta_4=\Delta_2$. The displacement of particles at the instant
of collision is $\Delta_{2,4}=D$.

To summarise this chapter, we have shown that the space-time and
the particle-hole symmetric forms of the non-local Boltzmann equation
are equivalent if the scattering rate includes the in-medium effect
on the level of the Bruckner reaction matrix.

\subsection{Comments on the Galitskii-Feynman T-matrix}
The Bruckner approximation of the scattering rate has been quite common
in earlier microscopic studies of the heavy ion reactions. Recently,
most studies prefers the Galitskii-Feynman approximation \cite{D84,BM90}
for which the Pauli blocking of the internal states is controlled by the
two-particle spectral function, ${\cal A}_{GF}=(G^>G^>)-(G^<G^<)$. 
Formula (\ref{AGalF}) is the quasiparticle approximation of 
${\cal A}_{GF}$.

As mentioned, the Galitskii-Feynman approximation includes processes
which go beyond the scope of naive kinetic equations. While terms 
$\propto (G^>G^>)$ exclude correlation in the occupied phase space, 
terms $\propto (G^<G^<)$ describe stimulated correlation by already 
existing pairs. These processes lead to the superconducting phase 
transition at low temperatures, therefore the system cannot be treated 
as a sum of single-particle excitations on the Fermi liquid ground 
state. Kinetic equations (\ref{Fbloc}) and (\ref{Fhole}) are based on 
the idea of the simple Fermi liquid and do not include the stimulated 
processes.

Even with the stimulated processes included, the kinetic equation can
be rearranged into the space-time symmetric form. Again, we make the
rearrangement on the level of the Green functions. Scattering integrals
(\ref{e7}) can be written so that their final states are consistent with
the Galitskii-Feynman spectral function,
\begin{eqnarray}
\{G^>,\Sigma^<\}\!-\!\{G^<,\Sigma^>\}&&=
\left\{G^>,G^>\circ T^R(G^<G^<)T^A\right\}
\nonumber\\
&&-\left\{G^<,G^<\circ T^R(G^<G^<)T^A\right\}
\nonumber\\
 -\bigl\{G^<,G^<\circ &&T^R((G^>G^>)-(G^<G^<))T^A\bigr\}.
\nonumber\\
\label{e7GF}
\end{eqnarray}
The last term includes the Galitskii-Feynman two-particle spectral
function and can be converted into the anti-causal picture. The
causal/anti-causal forms of the resulting kinetic equation reads,
\begin{eqnarray}
{\partial f_1\over\partial t}&+&{\partial\epsilon_1\over\partial k}
{\partial f_1\over\partial r}-{\partial\epsilon_1\over\partial r}
{\partial f_1\over\partial k}
\nonumber\\
&=&\int P^-\ f_3^-f_4^-\bigl(1-f_1-f_2^-\bigr)
\nonumber\\
&-&\int P^\mp\ \bigl(1-f_3^\mp -f_4^\mp\bigr)f_1f_2^\mp .
\label{e11GF}
\end{eqnarray}

The particle-hole symmetric form (with superscripts $-$) can be recast
into (\ref{Fhole}) by a subtraction of stimulated processes, $\propto
f_3^-f_4^-f_1 f_2^-$ on both sides. 
In contrast, the space-time symmetric form (with
superscripts $+$) cannot be recast to intuitive form (\ref{Fbloc}) due
to gradient contributions of stimulated processes, $\propto
f_3^+f_4^+f_1
f_2^+$. It is a pity, since the space-time symmetry is obligatory for
numerical treatments based on the Monte-Carlo simulations. Equation
(\ref{e11GF}) is not suited for the Monte-Carlo treatment because its
scattering integrals can change their sign loosing their probabilistic
interpretation. The Galitskii-Feynman type of kinetic equation
(\ref{e11GF}) thus provides a more precise description of the system,
but on cost of serious increase of difficulties of its numerical
treatment. Because of these numerical problems, we discuss an
implementation of symmetries only for the Bruckner approximation.

\subsection{Implementation of the symmetry in simulations}
Equivalency of both forms of kinetic equation, (\ref{e11ni}) and
(\ref{e11i}), offers an important simplification of the numerical
treatment. Expanding the scattering-out to the linear terms, one finds
that amplitudes of anti-causal and causal corrections equal while signs
are opposite. Since both forms are equivalent, the sum of gradient
corrections to the scattering-out vanishes.

For highly inhomogeneous and/or fast evolving systems, like the nuclear
matter in a heavy ion reaction, the Monte-Carlo simulation procedure
spends a majority of the CPU time searching when and where a collision
should be generated. Due to cancellation of gradient corrections to the
scattering-out integral, this part of the simulation procedure remains
the same as in the local approximation. All non-local corrections are
included only after the collision event is selected. This scheme has
been used in \cite{MSLKKN99}.

\section{Conclusions}
We have shown that the space-time symmetry of the non-local scattering
integral becomes non-trivial if the Pauli exclusion principle has to
be accounted for. Within the pseudo-classical form of the Pauli blocking
represented by the hole distributions as introduced by Nordheim, Uehling 
and Uhlenbeck, the space-time symmetry and the particle-hole symmetry
are consistent only if the scattering cross section includes in-medium
effects of Bruckner type. Due their classical form, these non-local
corrections are easily implemented into the Monte-Carlo simulations.

The more sophisticated approximation of Galitskii and Feynman includes
the stimulated creation of the colliding pair. This process escapes the
pseudo-classical interpretation of the scattering process what makes 
its implementation within the traditional Monte-Carlo simulation 
schemes impossible.

\acknowledgements
This work was supported from the GAASCR, Nr.~A1010806, the GACR,
Nos.~202/00/0643 and the Max-Planck-Society.

\appendix
\section{Optical theorem}
Identity (\ref{e8}) represents two alternative expressions of the
anti-hermitian part of the T-matrix,
\begin{equation}
M={\rm Im}T=i(T^R-T^A).
\label{a1}
\end{equation}
We derive this identity know as the optical theorem from the ladder
approximation, which is the approximation used in the scattering
integrals of the discussed kinetic equation.

The ladder approximation in the differential form reads,
\begin{equation}
T_{R,A}^{-1}=V-{\cal G}^{R,A}.
\label{a2}
\end{equation}
Here, ${\cal G}^{R,A}$ are the two-particle propagators given by the
time cut of the spectral function ${\cal A}=i({\cal G}^R-{\cal G}^A)$.
>From (\ref{a2}) follows
\begin{equation}
i\left(T_R^{-1}-T_A^{-1}\right)=-{\cal A}.
\label{a3}
\end{equation}
Multiplying (\ref{a1}) by $T_R^{-1}$ one finds
\begin{equation}
T_R^{-1}M=i-iT_R^{-1}T_A.
\label{a4}
\end{equation}
Finally, we express $T_R^{-1}$ from (\ref{a3}),
\begin{equation}
T_R^{-1}=T_A^{-1}+i{\cal A},
\label{a5}
\end{equation}
so that (\ref{a4}) turns into the familiar optical theorem
\begin{equation}
M=T^R{\cal A}T^A.
\label{a6}
\end{equation}

To obtain a less familiar anti-causal form of the optical theorem,
we multiply (\ref{a1}) by $T_R^{-1}$ from the right hand side,
\begin{equation}
MT_R^{-1}=i-iT_AT_R^{-1}.
\label{a7}
\end{equation}
Now we substitute (\ref{a5}) into (\ref{a7}) what yields
\begin{equation}
M=T^A{\cal A}T^R.
\label{a8}
\end{equation}
Comparing (\ref{a6}) with (\ref{a8}) one obtains identity (\ref{e8}).


\begin{thebibliography}{10}

\bibitem{E21}
D. Enskog,  in {\em Kinetic theory}, edited by S. Brush (Pergamon Press,
New
  York, 1972), Vol.~3, orig.: K. Svenska Vet. Akad. Handl. {\bf 63}(4)
(1921).

\bibitem{CC90}
S. Chapman and T.~G. Cowling, {\em The Mathematical Theory of
Non-uniform
  Gases} (Cambrigde University Press, Cambridge, 1990), third edition
Chap. 16.

\bibitem{C62}
E. Cohen, {\em Fundamental Problems in Statistical Mechanics}
(Nort-Holland,
  Amsterdam, 1962).

\bibitem{W65}
J. Weinstock, Phys. Rev. A {\bf 140},  460  (1965).

\bibitem{KO65}
K. Kawasaki and I. Oppenheim, Phys. Rev. A {\bf 139},  1763  (1965).

\bibitem{DC67}
J.~R. Dorfman and E.~G. Cohen, J. Math. Phys. {\bf 8},  282  (1967).

\bibitem{BE79}
H. van Beijeren and M.~H. Ernst, J. Stat. Phys. {\bf 21},  125  (1979).

\bibitem{B69}
K. B{\"a}rwinkel, Z. Naturforsch. a {\bf 24},  38  (1969).

\bibitem{La89}
F. Laloe, J. Phys. (Paris) {\bf 50},  1851  (1989).

\bibitem{L90}
D. Loos, J. Stat. Phys. {\bf 59},  691  (1990).

\bibitem{SN90}
R. Snider, J. Stat. Phys. {\bf 61},  443  (1990).

\bibitem{H90}
M. de~Haan, Physica A {\bf 164}, 373 (1990); {\bf 165} 224 (1990);
{\bf 170} 571 (1991).

\bibitem{NTL91}
P.~J. Nacher, G.~Tastevin and F.~Laloe,
Ann. Phys. (Leipzig) {\bf 48}, 149 (1991); J. Phys. (Paris) I~1, 181
(1991).

\bibitem{BKKS96}
T.~Bornath, D.~Kremp, W.~D.~Kraeft, and M.~Schlanges,
Phys. Rev. E {\bf 54}, 3274 (1996).

\bibitem{DP96}
P. Danielewicz and S. Pratt, Phys. Rev. C {\bf 53},  249  (1996).

\bibitem{SLM98}
V.~{\v S}pi{\v c}ka, P.~Lipavsk{\'y} and K.~Morawetz,
Phys. Lett. A {\bf 240}, 160 (1998).

\bibitem{Mt98}
P. Lipavsk{\'y}, K. Morawetz, and V. {\v S}pi{\v c}ka,
  Annales de Physique {\bf 26},1 (2001), K. Morawetz, Habilitation
University Rostock 1998.

\bibitem{N28}
L.~W. Nordheim, Proc. Roy. Soc. (A) {\bf 119},  689  (1928).

\bibitem{UU33}
E.~A. Uehling and G.~E.~Uhlenbeck, Phys. Rev. {\bf 43},  552  (1933).

\bibitem{Kel64}
L.~V. Keldysh, Zh. exper. teor. Fiz. {\bf 47},  1515  (1964).

\bibitem{LW72}
D.~C. Langreth and J.~W.~Wilkins, Phys. Rev. B {\bf 6},  3189  (1972).

\bibitem{BS94}
L. Banyai and K.~E.~Sayed, Ann. Phys. {\bf 233},  165  (1994).

\bibitem{GW64}
M.~L. Goldberger and K.~M.~Watson,
{\em Collision Theory} (Wiley, New York, 1964).

\bibitem{KB62}
L.~P. Kadanoff and G. Baym, {\em Quantum Statistical Mechanics}
(Benjamin, New
  York, 1962).

\bibitem{D84}
P. Danielewicz, Ann. Phys. (NY) {\bf 152},  239  (1984).

\bibitem{BM90}
W. Botermans and R. Malfliet, Phys.Rep. {\bf 198},  115  (1990).

\bibitem{LSM99}
P.~Lipavsk{\'y}, V.~{\v S}pi{\v c}ka and K.~Morawetz,
Phys. Rev. E {\bf 59}, R1291 (1999).

\bibitem{MSLKKN99}
K. Morawetz {\it et~al.}, Phys. Rev. Lett. {\bf 82},  3767  (1999).

\end{thebibliography}

\end{document}